\documentstyle[eqsecnum,prd,aps,epsf]{revtex}

\draft
\begin{document}
\newcommand{\beq}{\begin{equation}}
\newcommand{\eeq}{\end{equation}}
\newcommand{\beqn}{\begin{eqnarray}}
\newcommand{\eeqn}{\end{eqnarray}}
\newcommand{\pa}{\partial}
\newcommand{\vp}{\varphi}
\twocolumn[\hsize\textwidth\columnwidth\hsize\csname
@twocolumnfalse\endcsname

\begin{center}
{\large\bf{Innermost stable circular orbits around relativistic 
rotating stars}}
~\\
~\\
Masaru Shibata and Misao Sasaki\\
%\footnote{Email address: shibata~or~misao@vega.ess.sci.osaka-u.ac.jp} \\
{\em Department of Earth and Space Science,~Graduate School of
  Science,~Osaka University,\\
Toyonaka, Osaka 560-0043, Japan}\\
\end{center}

\begin{abstract}
We investigate the innermost stable circular orbit (ISCO) of 
a test particle moving on the equatorial plane around 
rotating relativistic stars such as neutron stars. 
First, we derive approximate analytic formulas for the angular velocity and 
circumferential radius at the ISCO making use of 
an approximate relativistic solution which is characterized by 
arbitrary 
mass, spin, mass quadrupole, current octapole and mass $2^4$-pole moments. 
Then, we show that the analytic formulas are accurate enough by 
comparing them with numerical results, which 
are obtained by analyzing the vacuum exterior around 
numerically computed geometries for rotating stars of polytropic equation 
of state. We demonstrate that contribution of mass quadrupole moment for 
determining the angular velocity and, in particular, 
the circumferential radius at the ISCO around a rapidly rotating star 
is as important as that of spin. 
\end{abstract}
\vskip2pc]

\baselineskip 5mm

\section{Introduction}

Observations of the low-mass X-ray binaries (LMXBs) with the 
Rossi X-ray Timing Explorer have revealed quasi-periodic 
oscillations (QPOs) around a frequency of $\sim 1$ kHz \cite{van}.  
At present, more than 10 sources 
of kHz QPOs have been found\cite{van}. 
One of the most impressive features of kHz QPOs is their 
very high frequency itself. 
The LMXBs are considered to be systems which include 
a neutron star of mass $M \sim 1.4M_{\odot}$, where 
$M_{\odot}$ denotes the solar mass, 
and the accretion disk around the neutron star. 
The Kepler frequency of the accretion disk is 
\beq
f \simeq 1.185{\rm kHz}\biggl({15{\rm km} \over R}\biggr)^{3/2}
\biggl({M \over 1.4M_{\odot}}\biggr)^{1/2}, 
\eeq
where $R$ denotes the circumferential radius around the neutron star. 
If the origin of the kHz QPOs is certain oscillation frequencies of 
the accretion disk surrounding a neutron star of low magnetic field, 
they must be generated at less than ten 
Schwarzschild radii of the neutron star. 
This means that the kHz QPOs may bring us a chance to explore 
general relativistic effects\cite{miller}. 

Several authors have recently suggested that 
at least some of the kHz QPOs may be related to 
the Kepler frequency at 
the innermost stable circular orbit (ISCO) of the accretion disk 
around a neutron star \cite{van}. 
One of the most strong reasons is that 
in many sources, the maximum frequency of the kHz QPOs is 
in narrow range between 1.1 and 1.2kHz, although they are thought to 
have very different mass accretion rates and 
magnetic fields. Due to the fact that 
the ISCO is determined by the property of the central neutron star, 
but not by the properties of the accretion disk, such as the mass 
accretion rate, it has been suggested that the origin of the kHz QPOs 
of the highest frequencies may 
be the Kepler frequency at the ISCO. 
If this is true, it means that we have a great opportunity to investigate 
general relativistic effects \cite{wagoner}\cite{miller}. 

Another remarkable feature of kHz QPOs is that they exhibit
some evidences that in the center of their sources, 
rapidly rotating neutron stars are involved as follows: 
(1) many sources display two peaks of 
the kHz QPO, and the frequency difference between the twin peaks $(\Delta f)$ 
changes only slightly with time 
although the frequency of each peak changes\cite{van}; 
(2) some sources which possess twin peaks also exhibit very coherent 
oscillations of several hundred Hz in X-ray bursts\cite{van}, and 
the frequencies change little with time during the bursts. 
Furthermore, they are approximately equal to or twice $\Delta f$. 
Since the spin frequency 
of a neutron star is the only candidate which changes only slightly 
on short time scales, 
the origin of the frequency difference between the twin peaks 
of QPOs and the oscillation frequencies in the X-ray bursts 
seem to correspond to the spin frequencies (or twice them) of 
neutron stars\cite{van}. 
This means that QPO sources include rapidly rotating neutron stars. 

Since the ISCO is determined by the geometry around the 
star, it is important to ask if the geometry around a neutron star 
can be approximated by that around a black hole. 
If the electric charge is neutral, stationary black holes must 
be of the Kerr type due to the uniqueness theorem \cite{hawking}. 
Kerr black holes have mass, spin, quadrupole 
moment and so on, but multipole moments higher than the 
quadrupole 
are expressed in terms of the mass $M$ and the spin angular momentum 
$J=S_1=qM^2~(|q| < 1)$ as $M_{2l}=M(iqM)^{2l}$ and 
$S_{2l-1}=-i M(iqM)^{2l-1}~(l=1,~2,~3,\cdot \cdot)$\footnote{
In this paper, we use units of $G=1=c$, where $G$ and $c$ are the 
gravitational constant and speed of light.} \cite{hansen}, 
where $M_{l}$ and $S_{l}$ denote the 
mass and current moments, respectively. 
This means that the geometry outside the 
black hole horizon is expressed only in terms of $M$ and $q$.
As a result, the ISCO is determined solely by them.
However, this is not the case for neutron stars. In the neutron star case, 
multipole moments higher than and including the quadrupole do not depend 
on the mass and spin in such a simple manner, and they are determined when 
the equation of state of the neutron star is given. 

The purpose of this paper is to point out the significance of the 
multipole moments (in particular, the mass quadrupole moment) 
of rapidly rotating neutron stars in determining the ISCO around them. 
This is due to the following fact: 
In the case of a Kerr black hole, the magnitude of the quadrupole moment  
is denoted by $M_2=-q^2M^3$\cite{hansen}, and $|M_2|$ is always 
smaller than $M^3$. 
In this case, the effect of the quadrupole moment is 
not important except in the case 
$q \sim 1$. However, in the case of a rotating neutron star, 
$|M_2/q^2M^3|$ may be much larger than unity $\sim 10$, and hence 
for rapidly rotating neutron stars which seem to be located at 
the center of QPO sources, $|M_2|$ may be larger than $qM^3$ 
for the case $q \agt 0.1$\cite{eric}. 
In such a case, the effect due to the quadrupole moment is as 
important as that due to the spin. 

The paper is organized as follows. 
In Sec. II, we derive approximate analytic formulas for 
the angular velocity and circumferential radius of the ISCO 
around a rotating object characterized by its mass, spin angular momentum, 
mass quadrupole, current octapole, and mass $2^4$-pole moments. 
In Sec. III, we perform numerical computations for obtaining 
stationary, axisymmetric 
spacetimes of rotating stars, and making use of the numerical results, 
we demonstrate that the accuracy of our formulas derived in Sec. II 
are accurate. Sec. IV is devoted to a summary. 

\section{Approximate analytic formulas for the angular velocity and 
circumferential radius at ISCO}

\subsection{Basic equations}

The line element of the vacuum exterior outside a stationary, 
axisymmetric rotating object is generally written as\cite{Wald}
\beq
ds^2=-F(dt-\omega d\varphi)^2+{1 \over F}\Bigl[
e^{2\gamma}(d\rho^2 +dz^2)+\rho^2 d\varphi^2 \Bigr]. 
\eeq
Throughout this paper, 
we assume that the spacetime has reflection symmetry with 
respect to the equatorial plane $z=0$, so that 
$F$, $\omega$, and $\gamma$ are functions of $\rho$ and $|z|$. 

Our purpose is to derive approximate formulas for 
the angular velocity and circumferential radius at the ISCO 
of a test particle on the equatorial plane around 
a rotating object. In the case when 
the test particle stays on the equatorial plane $(z=0)$, 
geodesic equations can be integrated to give
\beqn
{dt \over d\tau}&=&{E g_{\varphi\varphi} +\ell g_{t\varphi} \over g_2},\\
%%%%%%%%%%%%%%
{d\varphi \over d\tau}&=&{E g_{t\varphi} +\ell g_{tt} \over g_2},\\
%%%%%%%%%%%%%%
g_{\rho\rho}\biggl({d\rho \over d\tau}\biggr)^2&=&-1
+E^2{g_{\varphi\varphi} \over g_2}+2E\ell{g_{t\varphi} \over g_2}
+\ell^2{g_{tt} \over g_2} \nonumber \\
&\equiv& -V(\rho),
\eeqn
where $E$ and $\ell$ denote the specific energy and specific angular 
momentum of the test particle. $g_{tt}=-F$, $g_{t\vp}=F\omega$, 
$g_{\vp\vp}=-F\omega^2+\rho^2/F$, and 
$g_2 \equiv g_{t\varphi}^2-g_{tt}g_{\varphi\varphi}$ ($g_2=\rho^2$ in the 
present line element). 
For circular orbits, relations of the angular velocity 
$\Omega$, $E$, and $\ell$ are derived from 
the conditions $V=0=dV/d\rho$ as 
\beqn
&&\Omega \equiv {d\varphi \over dt} 
={-g_{t\varphi,\rho}+\sqrt{
(g_{t\varphi,\rho})^2-g_{tt,\rho}g_{\varphi\varphi,\rho}} 
\over g_{\varphi\varphi,\rho}}, \label{eqome}\\
%%%%%%%%%%%%%%%%%%%%%%%%%%%%%%
&& E=-{g_{tt}+g_{t\varphi}\Omega \over 
\sqrt{-g_{tt}-2g_{t\varphi}\Omega-g_{\varphi\varphi}\Omega^2} },
\label{eqene}\\
%%%%%%%%%%%%%%%%%%%%%%%%%%%%%%
&& \ell={g_{t\varphi}+g_{\varphi\varphi}\Omega \over
\sqrt{-g_{tt}-2g_{t\varphi}\Omega-g_{\varphi\varphi}\Omega^2} }. \label{eqang}
\eeqn
For simplicity,  we only consider prograde orbits in this paper. 
A circular orbit is stable (unstable) if 
\beq
{d^2V \over d\rho^2}={1 \over g_2}\biggl(
{d^2g_2 \over d\rho^2} -E^2{d^2g_{\varphi\varphi} \over d\rho^2}
 -2E\ell{d^2g_{t\varphi} \over d\rho^2}
 -\ell^2{d^2g_{tt} \over d\rho^2}
\biggr) \label{eqISCO}
\eeq
is positive (negative). Hence, the 
coordinate radius $\rho$ and $\Omega$ at the ISCO are determined from 
the condition where $d^2V/d\rho^2$ is vanishing. 

Note that Eq. (\ref{eqISCO}) is independent of 
metric function $\gamma$. 
Also, the angular velocity $\Omega$ 
and the circumferential radius ($\sqrt{g_{\vp\vp}}\equiv R$) are 
independent of $\gamma$. Thus, we only need $F$ and $\omega$ in the 
following. 

\subsection{Metric from the Ernst potential}

A stationary axisymmetric vacuum geometry is 
completely determined by the Ernst potential\cite{Ernst}, which is 
defined as 
\beq
{\cal E}=F+i\psi={\sqrt{\rho^2+z^2}-\xi \over \sqrt{\rho^2+z^2}+\xi},
\label{eqE}
\eeq
where $F=-g_{tt}$ and $\xi$ is a complex potential. 
If we know $\psi$, $\omega$ is calculated as 
\beq
\omega=-\int^{\rho}_{\infty} {\rho' \over F^2}{\pa \psi \over \pa z}d\rho' 
\Big|_{{\rm constant}~z}. \label{eqW}
\eeq
Thus, once we know $\xi$, we have all the necessary information. 

$\xi$ has the property that it can be expanded as \cite{FHP}\cite{Fintan}
\beq
\xi=\sum_{j=0}^{\infty}\sum_{k=0}^{\infty} a_{2j,k} {\rho^{2j} z^k \over 
(\rho^2 + z^2)^{2j+k}},
\eeq
where $a_{j,k}$ are complex constants in which information of 
the multipole moments of spacetime is completely contained. 
Note that $a_{j,k}$ is non-zero only for non-negative, even $j$ and 
non-negative $k$. Also, because of reflection symmetry with respect to the 
equatorial plane, $a_{j,k}$ is real for even $k$, and pure imaginary 
for odd $k$. Note that for investigation of the ISCO on the equatorial plane, 
we only need $a_{j,0}$ and $a_{j,1}$. 

Fodor, Hoenselaers, and Perjes (FHP)\cite{FHP} 
show that all the components of $a_{j,k}$ are derived 
by the following recursive relation
\beqn
a_{r,s+2}=&&{1 \over (s+1)(s+2)}\biggl[
-(r+2)^2a_{r+2,s} \nonumber \\
&&+\sum_{k,l,p,j}a_{k,l}a_{r-k-p,s-l-j}^* \times \nonumber \\
&&\{ a_{p,j}(p^2+j^2-4p-5j-2pk-2jl-2) \nonumber \\
&& +a_{p+2,j-2}(p+2)(p+2-2k) \nonumber \\
&& +a_{p-2,j+2}(j+2)(j+1-2l)\} \biggr],\label{receq}
\eeqn
where the sum is taken for $0 \leq k\leq r$, $0\leq l\leq s+1$, 
$0 \leq p\leq r-k$, and $-1 \leq j\leq s-l$\cite{Fintan}, and 
$a_{j,k}^*$ denotes the complex conjugate of $a_{j,k}$. 
As pointed out in ref.\cite{Fintan}, it can be shown that 
$a_{r,s+2}~(s \geq 0)$ is a function of $a_{j,0}$ and $a_{j-1,1}$ 
with $j\leq r+s+2$. This means that if we know $a_{j,0}$ and 
$a_{j,1}$ for $j \geq 0$, the entire spacetime metric (and 
of course, the metric on the equatorial plane) are completely determined. 
Note that 
if we know $a_{j,0}$ up to $j=2n$ and $a_{j,1}$ up to $j=2n-2$, we 
can calculate $a_{0,k}$ up to $k=2n$. Conversely, if 
we know $a_{0,k}$ up to $k=2n$, we can calculate 
$a_{j,0}$ up to $j=2n$ and $a_{j,1}$ up to $j=2n-2$. 

In principle, we can calculate terms of $a_{j,k}$ up to arbitrary large 
$j$ and $k$ using Eq. (\ref{receq}). In practice, however, we have to 
truncate higher terms. To access an appropriate truncation point, 
we can make use of the solution for slowly rotating black holes of 
the mass $M$ and the angular momentum $J(=S_1) \ll M^2$. 
In this case, 
%components of spacetime metric around the equatorial plane are
%\beqn
%&& g_{tt}=-{\sqrt{\rho^2+M^2} - M \over  \sqrt{\rho^2+M^2} +M}+O(z^2),\\
%&& g_{\vp\vp}=-{1 \over g_{tt}}+O(z^2),\\
%&& g_{t\vp}=-{2J \over\rho^2}\Bigl(\sqrt{\rho^2+M^2} - M \Bigr)+O(z^2), 
%\eeqn
%and $\xi$ around the equatorial plane becomes 
\beq
\xi={M \rho \over \sqrt{\rho^2+M^2}}+{iz\rho J \over (\rho^2+M^2)^{3/2}}+
O(z^2),
\eeq
where we have neglected terms of $O(J^2/M^4)$. 
Note that $\rho$ is related to the Schwarzschild radial 
coordinate $r_s$ as $\rho=\sqrt{r_s(r_s-2M)}$. The 
Schwarzschild coordinate radius of the ISCO is $r_s=6M(1-\sqrt{8/27}J/M^2)$, 
so that the radius of the ISCO in $\rho$ is given by 
\beq
\rho_{\rm ISCO}=\sqrt{24}M-{10J \over 3M}. 
\eeq 
We have investigated how the estimated values of 
$\rho_{\rm ISCO}$ differ from the above one by 
expanding $\xi$ in terms of $O(M/\rho)$ as 
\beqn
\xi&=&1-{M^2 \over 2\rho^2}+{3M^4 \over 8\rho^4}
-{5M^6 \over 16\rho^6}+{35M^8 \over 128\rho^8} 
-{63M^{10} \over 256\rho^{10}} \nonumber \\
&&+i{z J  \over \rho^2}\biggl(1-{3M^2 \over 2\rho^2}
+{15 M^4 \over 8\rho^4}-{35M^6 \over 16\rho^6}+
{315 M^8 \over 128 \rho^8} \biggr)\nonumber \\
&&+O(\rho^{-12}). 
\eeqn
We have found that in the case when we take up to 
$O(\rho^{-6})$,  $O(\rho^{-8})$, and  $O(\rho^{-10})$ terms, 
the errors in $\rho_{\rm ISCO}$ are 
about $4\times 10^{-4}$, $10^{-5}$ and $4\times 10^{-7}$, 
respectively, for the coefficient of $M$, and 
about $2\times 10^{-3}$, $10^{-4}$ and $4\times 10^{-6}$, 
respectively, for the coefficient of $J/M$. 
The order of magnitude of the errors in 
$\Omega_{\rm ISCO}$ have turned out to be almost the same as those 
in $\rho_{\rm ISCO}$. 
Thus, we expect 
that this method will generate a fairly accurate approximate formula 
even if we include the higher multipole moments. 
In the following, we take $a_{j,0}$ up to $j=10$ and 
$a_{j,1}$ up to $j=8$, i.e., we take into account all the terms 
up to $O(\rho^{-10})$ consistently. In other word, we calculate 
$a_{0,k}$ up to $k=10$, and neglect $a_{0,k}$ of $k\geq 11$.

\subsection{Results}

Our strategy for determining the ISCO is as follows. 
First, we assume that the spacetime is characterized by 
mass $M$, spin angular momentum $J(=S_1)=qM^2$, 
mass quadrupole $M_2=-Q_2M^3$, 
current octapole moments $S_3 = -q_3M^4$, and mass $2^4$-pole $M_4=Q_4M^5$, 
and neglect the higher multipole moments. Note that 
$q$, $Q_2$, $q_3$, and $Q_4$ are positive for a rotating star 
in a prograde spin.

In the case of a rotating neutron star, 
we may assume $q \sim O(\epsilon_1)$, 
$Q_2 \sim O(\epsilon_2)$, $q_3 \sim O(\epsilon_1 \epsilon_2)$, 
$Q_4 \sim O(\epsilon_2^2)$, where 
$\epsilon_1,~\epsilon_2 \ll 1$, because 
$q$ and $Q_2$ are expected to be less than unity. 
For a slowly rotating neutron star, $Q_2 \sim O(q^2)$, 
so that $\epsilon_2 \sim O(\epsilon_1^2)$. 
In this case, $Q_2 \sim O(\epsilon_1^2)$, $q_3 \sim O(\epsilon_1^3)$,
$Q_4 \sim O(\epsilon_1^4)$. 
However, for a rapidly rotating neutron star, $Q_2$ can be as large as $q$, 
so that whenever the terms proportional to $q^2$ make a significant 
contribution, we should 
also take into account the terms proportional to $Q_2^2$, $Q_4$, $q q_3$, 
and so on. This is the reason why we take into account $S_3 $ and $M_4$. 

Hereafter, we expand all the quantities by means of 
$\epsilon \equiv \epsilon_1$ by formally setting 
$q \rightarrow \epsilon q$, 
$Q_2 \rightarrow \epsilon^2 Q_2$,$q_3 \rightarrow \epsilon^3 q_3$,
and $Q_4 \rightarrow \epsilon^4 Q_4$, and retain 
all the terms up to $O(\epsilon^4)$ consistently. 
Thus, the formulas derived below are accurate up to $O(\epsilon_1^4)$ 
for a slowly rotating neutron star. 
Even for a rapidly rotating neutron star, the formulas include all 
the terms of $O(\epsilon_1)$, $O(\epsilon_2)$, 
$O(\epsilon_1^2)$, $O(\epsilon_2^2)$, and $O(\epsilon_1\epsilon_2)$. 
Hence, they are still accurate to $O(Q_2^2)$. 

As mentioned in the previous subsection, 
we need $a_{0,k}$ up to $k=10$. Relation between 
multipole moments and $a_{0,k}$ for $0 \leq k \leq 10$ have 
been already given by FHP\cite{FHP}. 
Hence, by using them, we 
can calculate $a_{2j,0}~(1 \leq j \leq 5)$ and 
$a_{2j,1}~(1 \leq j \leq 4)$ up to $O(\epsilon^4)$ using 
Eq. (\ref{receq})\cite{mathe}. 
(The explicit forms of $a_{0,k}$, $a_{2j,0}$ and $a_{2j,1}$ 
are shown in Appendix A). 
Once $\xi$ is determined up to $O(\rho^{-10})$, 
$F$ and $\omega$ are straightforwardly 
obtained from Eqs. (\ref{eqE}) and (\ref{eqW}) in the following form;
\beqn
&&F=1+\sum_{j=1}^{11} C_{F,j} \biggl({M \over \rho}\biggr)^j +O(\rho^{-12}),\\
&&\omega=\sum_{j=3}^{11} C_{\omega,j} {M^{j-1} \over \rho^j} +O(\rho^{-12}),
\eeqn
where $C_{F,j}$ and $C_{\omega,j}$ are functions of 
$q$, $Q_2$, $q_3$, and $Q_4$. 
Substituting these equations into $d^2 V/d\rho^2=0$, 
we rewrite Eq. (\ref{eqISCO}) in the form 
\beq
\sum_{i=0}^4 \epsilon^i A_i(\rho)=0, \label{eqalg}
\eeq
where the coefficients $A_i$ are independent of $\epsilon$ but 
depend on $q$, $Q_2$, $q_3$, $Q_4$, and $\rho$. 
These coefficients can be explicitly written out by substituting 
Eqs. (2.16) and (2.17) into Eq. (\ref{eqISCO}) 
(making use of Eqs. (\ref{eqene}) and (\ref{eqang})) and gathering 
in powers of $\epsilon$. 
We then look for a solution in the form 
\beq
\rho=\sum_{i=0}^4 \epsilon^i c_i ,
\eeq
where we know that $c_0=\sqrt{24}M$ and $c_1=-10qM/3$. 
Imposing that Eq. (\ref{eqalg}) holds in each order of $\epsilon$, 
it is rewritten into three algebraic equations 
for $c_2$, $c_3$, and $c_4$ as  
\beqn
&&{1 \over 2} {d^2 A_0 \over d\rho^2}c_1^2 + {d A_0 \over d\rho}c_2 
+{d A_1 \over d\rho}c_1 + A_2=0 ,\\
&&{1 \over 6} {d^3 A_0 \over d\rho^3}c_1^3 + {d^2 A_0 \over d\rho^2}c_1 c_2 
+ {d A_0 \over d\rho}c_3 \nonumber \\
&&~~~~~+{1 \over 2} {d^2 A_1 \over d\rho^2}c_1^2 + {d A_1 \over d\rho}c_2
+{d A_2 \over d\rho}c_1 + A_3=0 ,\\
&&{1 \over 24}{d^4A_0 \over d\rho^4}c_1^4
+{1 \over 2}{d^3A_0 \over d\rho^3}c_1^2c_2
+{1 \over 2}{d^2A_0 \over d\rho^2}(c_2^2+2c_1c_3) \nonumber \\
&&~~~~~+{dA_0 \over d\rho}c_4
+{1 \over 6} {d^3 A_1 \over d\rho^3}c_1^3 + {d^2 A_1 \over d\rho^2}c_1 c_2
+ {d A_1 \over d\rho}c_3 \nonumber \\
&&~~~~~+{1 \over 2} {d^2 A_2 \over d\rho^2}c_1^2 + {d A_2 \over d\rho}c_2
+{d A_3 \over d\rho}c_1 + A_4=0 ,
\eeqn
where $A_k$ and their derivatives are evaluated at $\rho=c_0$.
{}From these algebraic equations we obtain
\beqn
&&c_2\simeq (-1.41671q^2 + 1.07648 Q_2)M,\nonumber \\
&&c_3\simeq (-1.45081q^3 + 1.60313 qQ_2 - 0.32561q_3)M, \nonumber \\
&&c_4\simeq (-1.87941q^4 + 2.74953 q^2Q_2 - 0.37568 Q_2^2 \nonumber \\
&&~~~~~~~~~~~~+0.09377Q_4 - 0.70086 q q_3)M .
\eeqn
Using these approximate solutions, we reach 
approximate expressions for the angular velocity and 
circumferential radius of the ISCO, $\Omega_{\rm ISCO}$ and $R_{\rm ISCO}$, 
as follows. 
\beqn
&&\Omega_{\rm ISCO}={1 \over 6\sqrt{6}M}\Bigl(
1 +0.74846q +0.78059 q^2 -0.23429 Q_2 \nonumber \\
&&~~~~~~~~~~~~~~~~~+0.98094 q^3 - 0.64406 q Q_2 + 0.07432q_3 \nonumber \\
&&~~~~~~~~~~~~~~~~~+1.38118 q^4-1.41729 q^2Q_2 + 0.12798 Q_2^2 \nonumber \\
&&~~~~~~~~~~~~~~~~~-0.02129 Q_4 + 0.25028q q_3 \Bigr),\label{ISCOW} \\
%%%%%%%%%%%%%%%%%
&&R_{\rm ISCO}=6M\Bigl(
1-0.54433q - 0.22619q^2 + 0.17989Q_2 \nonumber \\
&&~~~~~~~~~~~~~~~~-0.23002 q^3 + 0.26296 q Q_2 - 0.05317 q_3 \nonumber \\
&&~~~~~~~~~~~~~~~~-0.29693 q^4 + 0.44546 q^2Q_2 -0.06249 Q_2^2 \nonumber \\
&&~~~~~~~~~~~~~~~~+0.01544 Q_4 - 0.11310 q q_3\Bigr).\label{ISCOR}
\eeqn
For the case of Kerr metric, $Q_2=q^2$, $q_3=q^3$, 
and $Q_4=q^4$, we obtain
\beqn
&&\Omega_{\rm ISCO}={1 \over 6\sqrt{6}M}\Bigl(
1 + 0.74846q + 0.54630 q^2 + 0.41120 q^3\nonumber \\
&&~~~~~~~~~~~~~~~~~~~~~~~~~~+0.32085 q^4 \Bigr),\\
&&R_{\rm ISCO}=6M\Bigl(
1 - 0.54433 q - 0.04630 q^2 -0.02023 q^3 \nonumber \\
&&~~~~~~~~~~~~~~~~~~~~~~~~~~-0.01162 q^4\Bigr).
\eeqn
On the other hand, the exact Kerr solution gives \cite{COMPACT} 
\beqn
&&\Omega_{\rm ISCO}^{\rm Kerr}={1 \over 6\sqrt{6}M}\Bigl(
1+0.74846q+0.54630q^2+0.41108q^3\nonumber \\
&&~~~~~~~~~~~~~~~~~~~~~+0.31991 q^4 +O(q^5)\Bigr),\\
&&R_{\rm ISCO}^{\rm Kerr}=6M\Bigl(
1-0.54433q - 0.04630q^2 -0.02016q^3\nonumber \\
&&~~~~~~~~~~~~~~~~~~~~~- 0.01110q^4+O(q^5)\Bigr).
\eeqn
Thus, the error of the coefficients 
in our approximate formulas of $\Omega_{\rm ISCO}$ and $R_{\rm ISCO}$ 
is less than  $10^{-4}$ for $O(\epsilon^2)$ terms, less than $10^{-3}$ for 
$O(\epsilon^3)$ terms, and $\sim 10^{-2}$ for $O(\epsilon^4)$ terms. 
Hence, for slowly rotating neutron stars ($q < 0.1 $), our formulas 
are accurate enough. Even for very rapidly rotating neutron stars 
of $q, Q_2 \sim 0.5$, 
we may expect that they will yield very accurate values. 

We note that in Eqs. (\ref{ISCOW}) and (\ref{ISCOR}), 
the signs of coefficients of the terms including $Q_2$ such as 
$Q_2$, $q Q_2$ and $q^2 Q_2$ 
are opposite to those of $q^k~(k=1-4)$ terms. As shown in ref.\cite{eric}, 
$Q_2/q^2 $ is always larger than unity for a rotating neutron star, and 
hence $Q_2$ may be of $O(q)$ for rapidly rotating neutron stars of 
$q \agt 0.1$. 
This implies that for rapidly rotating neutron stars, the effect due to 
the $q^n$ terms is significantly suppressed by that due to $Q_2$. 
Thus, we conclude that 
the effect of the quadrupole is important in 
determining the ISCO even when $q \sim 0.1$. 
On the other hand, the coefficients of $q_3$ and $Q_4$ are smaller than 
those of $q$ and $Q_2$. 
Thus, their contributions are small. 

The formulas of $E$ and $\ell$ at the ISCO are shown in Appendix B. 

\section{Numerical study}

To confirm the accuracy of the formulas derived in Sec. II, 
we compare them with numerical solutions. 
We numerically construct stationary, axisymmetric spacetimes of 
relativistic rotating stars with a polytropic equation of state using 
Komatsu, Eriguchi, and Hachisu method\cite{KEH}. 
Then, we estimate $\Omega_{\rm ISCO}$ and $R_{\rm ISCO}$ 
as well as the multipole moments of the numerically 
generated spacetimes. 
It is desirable to include all the multipole moments 
up to $Q_4$, and compare our analytic 
formulas with the numerical results. However, accurate 
numerical calculation of $Q_4$ is difficult. In this section, 
we simply set $Q_4= \alpha Q_2^4$. For the Newtonian 
incompressible case (Maclaurin spheroid case), $\alpha=15/7$, and 
for Newtonian compressible case, we find $\alpha < 15/7$. 
Due to general relativistic effects, stars are more centrally condensed 
than those in the Newtonian case, so that 
we expect that $\alpha$ in general 
relativity is smaller than that in the Newtonian theory, i.e., 
$0 <\alpha \alt 2$. Hence, we simply set $\alpha=1$. 
Fortunately, the coefficients of $Q_4$ in 
$\Omega_{\rm ISCO}$ and $R_{\rm ISCO}$ are small, so that our 
rough treatment does not affect the result much. 
In fact, we also set $\alpha=0$ and $2$, but 
our conclusion is not changed at all. 
 
\subsection{Basic equations}

We consider the energy momentum tensor of an ideal fluid, 
\beq
T_{\mu\sigma}=\rho_b\biggl(1+\varepsilon + {P \over \rho_b}\biggr)
u_{\mu}u_{\sigma} + Pg_{\mu\sigma},
\eeq
where $\rho_b$, $P$, $\varepsilon$, $u^{\mu}$, and $g_{\mu\sigma}$ are 
the baryon rest 
mass density, pressure, specific internal energy, four velocity, 
and spacetime metric. We adopt the polytropic equation of state, 
\beq
P=K \rho_b^{\Gamma}=(\Gamma-1)\rho_b\varepsilon ;~\Gamma=1+{1 \over n},
\eeq
where $K$ and $n$ are the polytropic constant and the polytropic index. 

We only consider the case when stars uniformly rotate around the $z$-axis. 
Hence, we set $u^r=u^{\theta}=0$ and 
$u^{\vp}=u^0 \Omega_s$, where $\Omega_s$ is the 
spin angular velocity of the rotating star. In this case, 
the fluid equations of motion are easily integrated to give\cite{LPPT}
\beqn
{h^2 \over (u^0)^2}&=&
-h^2(g_{tt}+2g_{t\vp}\Omega_s+g_{\vp\vp}\Omega_s^2 )\nonumber \\
&=&{\rm constant},\label{beleq}
\eeqn
where $h=1+\varepsilon+P/\rho_b=1+\Gamma \varepsilon$. 

Following Butterworth and Ipser \cite{BI}, we write the line element as 
\beqn
ds^2&=&-e^{2\nu}dt^2+r^2\sin^2\theta B^2 e^{-2\nu}(d\vp-\omega dt)^2
\nonumber \\
&&+e^{2\zeta-2\nu}(dr^2+r^2d\theta^2),
\eeqn
where $\nu$, $B$, $\zeta$, and $\omega$ are field functions depending on 
$r$ and $\theta$. We present the field equations for these variables 
and briefly describe our numerical method in Appendix C. 

Once the field variables are computed, the 
ISCO on the equatorial plane 
is found making use of Eq. (\ref{eqISCO}) 
together with Eqs. (\ref{eqome})$-$(\ref{eqang}). 
%Replacing $\rho$ into $r$, and noting $g_{\vp\vp}=r^2B^2e^{-2\nu}$, 
%$g_{t\vp}=-\omega g_{\vp\vp}$, $g_{tt}=-e^{2\nu}+\omega^2g_{\vp\vp}$, 
%and $g_2=g_{\vp\vp}e^{2\nu}$, the condition $d^2V/dr^2=0$ is written as
%\beqn
%&&g_{\vp\vp,rr}(\Omega-\omega)^2-4\omega_{,r}(\Omega-\omega)g_{\vp\vp,r}
%-2(\Omega-\omega)\omega_{,rr}g_{\vp\vp}\nonumber \\
%&&~-(e^{2\nu})_{,rr} +2(\omega_{,r})^2g_{\vp\vp}^2(\Omega-\omega)^2e^{-2\nu}
%\nonumber \\
%&&~-2(e^{2\nu})_{,r}(\ln g_{\vp\vp})_{,r}+4\nu_{,r}g_{\vp\vp,r}
%(\Omega-\omega)^2=0, 
%\eeqn
%where 
%\beq
%\Omega={-\omega_{,r}g_{\vp\vp}-\omega g_{\vp\vp,r}+
%\sqrt{(\omega_{,r})^2g_{\vp\vp}^2+(e^{2\nu})_{,r}g_{\vp\vp,r}} 
%\over g_{\vp\vp,r} }. 
%\eeq
To examine the accuracy of our approximate 
analytic formulas derived in Sec. II, we also 
need to estimate the multipole moments. 
For $M$ and $q$, we have the formulas as \cite{BI}
\beqn
&&M=\int d^3x \biggl[e^{2\zeta} \rho_b B \Bigl[2h(u^0)^2 - 
e^{-2\nu}\{1+\varepsilon(2-\Gamma)\} \Bigr] \nonumber \\
&&~~~~~~~~~~+2r^2\sin^2\theta (\Omega_s-\omega)\omega B^3 
(u^0)^2e^{2\zeta-4\nu}\rho_b h\biggr],\label{Meq} \\
&& q={1 \over M^2} \int d^3x r^2\sin^2\theta (\Omega_s-\omega)B^3
(u^0)^2e^{2\zeta-4\nu}\rho_b h. \label{Jeq}
\eeqn 
For $Q_2$ and $q_3$, we estimate them by using the 
asymptotic behaviors of $\nu$ and $\omega$. 
For $r\rightarrow \infty$, $\nu$, $\omega$ and $B$ behaves as \cite{BI}
\beqn
&&\nu\rightarrow 
-{M \over r}+{B_0M \over 3r^3}+{Q_2M^3 \over r^3}P_2(y) +O(r^{-4}),
\label{boundM}\\
&&\omega \rightarrow 
{2qM^2 \over r^3}-{6qM^3 \over r^4}+{6 \over 5}
\biggl[8-{3B_0 \over M^2}\biggr]{qM^4 \over r^5} \nonumber \\
&&~~~~~~~~~-{q_3 M^4 \over r^5}(5y^2-1)+O(r^{-6}),\label{boundw} \\
&&B \rightarrow 1+ {B_0 \over r^2}+O(r^{-4}),\label{boundB}
\eeqn
where $y=\cos\theta$, $P_2(y)=(3y^2-1)/2$, 
and $B_0$ is a constant which is $-M^2/4$ for 
spherical cases. 
When we numerically obtain $\nu$, $\omega$, and $B$, 
we can extract information of $Q_2$ and $q_3$ at a large radius $r \gg M$ 
near the outer numerical boundary ($r=r_{\rm max}$) as 
\beqn
&&Q_2=5 {r^3 \over M^3}\int_0^1 dy P_2(y) \nu ,\\
&&q_3=-{21 r^5 \over 16M^4} \int_0^1 dy  (5y^2-1)(1-y^2)\omega .
\eeqn
We evaluate $Q_2$ and $q_3$ at various large radii, and 
find that $Q_2$ quickly converges as $r \rightarrow r_{\rm max}$. 
This suggests that the error for estimation of $Q_2$ 
is very small (we guess it is less than $1\%$). 
This is mainly because the coefficient of $O(r^{-4})$ part 
in $\nu$ is not large. On the other hand, convergence of  
$q_3$ at  $r \rightarrow r_{\rm max}$ is slow 
because the coefficient of $O(r^{-6})$ part in $\omega$ is large \cite{BI}. 
To obtain $q_3$ accurately, it is desirable to attach the outer 
numerical boundary as $r_{\rm max} \gg M$, but 
to do that we need to take many grid points and 
hence need a long computational time. To save computational time, 
we use here an extrapolation method to estimate $q_3$; i.e., 
we calculate $q_3$ at several large radii which are not near the outer 
boundary ($r \sim 3r_{\rm max}/4$), and 
extrapolate true value of $q_3$ at $r \rightarrow \infty$ by assuming 
that $q_3$ behaves as $q_3(r)=q_3(\infty)+C/r$ for the large 
radii, where $C$ is a constant. Since this method is rough, 
we guess that the error of $q_3$ may be $\sim 10\%$ in this 
method. However, the large error in 
$q_3$ does not affect the following analysis much 
because the coefficients of $q_3$ terms 
in $\Omega_{\rm ISCO}$ and $R_{\rm ISCO}$ are 
fortunately small. The important quantities in our analysis 
are $M$, $q$ and $Q_2$. 
We mention that we have also evaluated $M$ and $q$ by using 
\beqn
&&M=-r\int_0^1 dy \nu ,\label{Meq2} \\
&&q=\biggl[1-{3M \over r}+{3 \over 5}\biggl(8-{3B_0 \over M^2}\biggr)
\biggr]^{-1}{3r^3 \over 4M^2} \nonumber \\
&&~~~~~~~~~~~~~~~~~~~~~\times \int_0^1 dy (1-y^2)\omega, \label{Jeq2}
\eeqn
and confirmed that $M$ calculated by Eqs. (\ref{Meq}) and (\ref{Meq2}) 
and $J$ calculated by Eqs. (\ref{Jeq}) and (\ref{Jeq2}), respectively, 
agree very well.

\subsection{Results}

As the polytropic index, we set $n=1$. 
In Fig. 1, we show the relation between $\rho_{b,0}$ ($\rho_b$ at 
$r=0$) and $M$ for the case $n=1$. 
Note that in the figure, we plot non-dimensional quantities 
$\rho_{b,0}K^n$ and $MK^{-n/2}$. 
The lower and upper solid lines denote 
the relations for the spherical star and for the rotating star 
at the mass shed limit, respectively. 
The filled squares denote data sets that are used for comparison with
our analytic formulas.
The dotted line is the critical line above which the ISCO ceases to
exist. For sufficiently large $\rho_{b,0}$, stars are unstable 
against radial gravitational collapse \cite{CST}. 
The dashed line divides the stable and unstable branches:
The left-hand side of it is the stable region. 
(Here, for judging the stability against the 
gravitational collapse, we have applied the turning point method 
shown in ref.\cite{FIS}.) 
Since unstable stars are not realized in nature, 
we exclude numerical data in the unstable region. 

In Table I, we show all the data we use in our analysis. 
As pointed out in ref.\cite{eric}, 
$Q_2/q^2$ is always greater than unity. 
(For the data in Table I, $2 \alt  Q_2/q^2 \alt 4$. )

In Fig.~2, we show 
\beqn
&&\Delta_{\Omega}(l) = 6\sqrt{6}M
\biggl( \Omega^{\rm numerical}_{\rm ISCO}
-\Omega_{\rm ISCO}(l)\biggr),\nonumber \\
&&\Delta_r (l)= 
{1 \over 6M}\biggl(R^{\rm numerical}_{\rm ISCO}
-R_{\rm ISCO}(l) \biggr), \label{delta}
\eeqn
as a function of $Q_2$. Here, $\Omega_{\rm ISCO}(l)$
and $R_{\rm ISCO}(l)~(1 \leq l \leq 4)$ denote the analytic formulas 
in which we include terms up to $O(\epsilon^l)$. 
We also define $\Delta_{\Omega}^{\rm Kerr}$ and 
$\Delta_r^{\rm Kerr}$ in which we use the analytic relations for the 
Kerr metric. The open circles, crosses, filled circles, open triangles
and open squares denote 
$\Delta_{a}^{\rm Kerr}$, $\Delta_{a}(1)$, 
$\Delta_{a}(2)$, $\Delta_{a}(3)$ and 
$\Delta_{a}(4)$~($a=\Omega~{\rm or}~r$), respectively. 

{}From Fig.2, we find 
apparently that the Kerr formulas for the ISCO are not appropriate at all. 
The formula $R_{\rm ISCO}(1)$ is as bad as the Kerr formula, but
$\Omega_{\rm ISCO}(1)$ is fairly good. 
The latter feature seems accidental: Since the relation 
$Q_2 \sim \alpha q^2$ with $\alpha \sim 2-4$ holds, 
cancellations between $q^2$ and $Q_2$ terms, between 
$q^3$ and $qQ_2$ terms, and among $q^4$, $q^2Q_2$ and $Q_4$ terms 
occur. (In contrast, if $Q_2 \sim q^2$, $R_{\rm ISCO}(1)$ will be bad, 
while $\Omega_{\rm ISCO}(1)$ will be good. 
If $Q_2 \sim 10q^2$, both formulas will not be good at all). 
%Besides the linear formulas, the $(l+1)$th-order formulas are always 
%better than the $l$th-order formulas. 

It should be stressed that for small $Q_2$, $\Delta_{\Omega}^{\rm Kerr}$ 
and $\Delta_r^{\rm Kerr}$ (also $\Delta_r(1)$) are 
roughly proportional to $Q_2$, while 
$\Delta_{\Omega}(2)$ and $\Delta_r(2)$ remain very small.
Recalling that in the Kerr formulas
$\Omega_{\rm ISCO}^{\rm Kerr}$ and $R_{\rm ISCO}^{\rm Kerr}$, 
the effect of $Q_2$ as an independent variable is absent,
this feature implies a substantial effect of $Q_2$ on the
determination of the ISCO around neutron stars even for $q\sim 0.1$. 

We find the errors of the formulas $\Omega_{\rm ISCO}(2)$ and 
$R_{\rm ISCO}(2)$ are always very small for small $Q_2 < 0.2$ 
($q\alt 0.2$ in this case). 
For large $Q_2 > 0.2$, however, the errors gradually increase. This
indicates that the effect of the terms of $O(Q_2^2)$ (such as $Q_4$)
is not negligible. Therefore, it should be necessary to correctly take
into account the terms of $O(Q_2^2)$, in order to give an appropriate 
formula for very rapidly rotating neutron stars. 
Note, however, that Fig. 2 also indicates that 
$\Delta_a~(a=\Omega~{\rm or}~r)$ adequately converges to zero by
adding higher order terms in $\epsilon$. 
Thus, unless $q,~Q_2 \agt 1$, 
an appropriate formula for very rapidly rotating neutron stars of 
large $q$ and $Q_2$ can be derived along the line 
presented in this paper. 

\section{Summary}

In this paper, we have investigated the ISCO of a test particle 
moving on the equatorial plane around a rotating object. 
We have derived fairly accurate approximate formulas for 
the angular velocity and circumferential radius of the 
ISCO including mass, spin, quadrupole, current octapole, mass $2^4$-pole 
moments of a rotating object. 
Our formulas show that the effect of quadrupole moment is 
important for determining the angular velocity and, in particular, the 
circumferential radius of the ISCO for 
rapidly rotating neutron stars of $Q_2 \agt q$\cite{eric}. 

In a recent paper, Miller, Lamb, and Cook \cite{miller} performed a 
numerical computation for analyzing the ISCO around 
rapidly rotating neutron stars, and pointed out 
that $\Omega_{\rm ISCO}$ and in particular $R_{\rm ISCO}$ 
are not correctly determined by the first 
order analytic treatment in $q$. 
Our present study clarifies the reason for 
inaccuracy of the first order formula which is mainly 
due to the neglection of the quadrupole moment.

\acknowledgments

Numerical computation was in part performed using FACOM VX/4R 
in the data processing center of National Astronomical Observatory 
of Japan (NAOJ). This work was in part supported by a Japanese 
Grant-in-Aid of Ministry of Education, Culture, Science and Sports 
(Nos. 08NP0801 and 09740336). 

\appendix
\section{Coefficients for the Ernst potential}

FHP \cite{FHP} give 
\beqn
&&a_{0,0}=M, ~~~a_{0,1}=i J, ~~~a_{0,2}=M_2, \nonumber \\
&&a_{0,3}=i S_3,~~~a_{0,4}=M_4+{M_{20}M \over 7}, \nonumber \\
%%%%%%%%%%%%%%%%%%%%%%%%
&&a_{0,5}=-i {JM_{20} \over 21}+{M_{30}M \over 3}, \nonumber \\
%%%%%%%%%%%%%%%%%%%%%%%%
&&a_{0,6}=-{M_{20}M^3 \over 33}+{5M_{20}M_2 \over 231}
-{4i M_{30}J \over 33}+{8 M_{31} M \over 33} \nonumber \\
&& \hskip 2cm +{6M_{40} M \over 11},\nonumber \\
%%%%%%%%%%%%%%%%%%%%%%%%
&& a_{0,7}={19 i M_{20}M^2 J \over 429}-{15 M_{30}M^3 \over 143} 
+M^8 O(\epsilon^5) , \nonumber \\
%%%%%%%%%%%%%%%%%%%%%%%%
&&a_{0,8}={M_{20}M^5 \over 143}+{53 M_{20}MJ^2 \over 3003}
-{311M_{20}M_2M^2 \over 3003} \nonumber \\
&&~~~~~+{36 i M_{30}M^2J \over 143}
-{3M^3 \over 13}\Bigl(M_{31}+M_{40}\Bigr)+M^9 O(\epsilon^5),
\nonumber \\
%%%%%%%%%%%%%%%%%%%%%%%%
&&a_{0,9}=-{43i M_{20}M^4J \over 2431}+{7M_{30}M^5 \over 221}
+M^{10}O(\epsilon^5),\nonumber \\
%%%%%%%%%%%%%%%%%%%%%%%%
&& a_{0,10}=-{7M_{20} M^7 \over 4199}+{4423M_{20}M^4 M_2  \over 138567}
-{202 M_{20}M^3 J^2 \over 12597} \nonumber \\
&&~~~~~-{462 i M_{30}M^4 J \over 4199}
+{28 M^5 \over 323}\Bigl(M_{31}+M_{40}\Bigr) \nonumber \\
&&~~~~~+M^{11} O(\epsilon^5),
\eeqn
where $M_{20}=MM_2+J^2$, $M_{30}=i(S_3M-M_2J)$, $M_{31}=-S_3J-M_2^2$, 
and $M_{40}=M_4M+S_3J$\cite{FHP}. 

Using Eq. (\ref{receq}), $a_{2j,0}~(1 \leq j \leq 5)$ and 
$a_{2j,1}~(1 \leq j \leq 4)$ up to $O(\epsilon^4)$ 
are calculated as 
\beqn
&&a_{2,0}=-{M^3 \over 2}(1-Q_2),\nonumber \\
&&a_{4,0}={M^5 \over 56}\Bigl(21+10q^2 - 52Q_2+21Q_4\Bigr),\nonumber \\
&&a_{6,0}=M^7\biggl(-{5 \over 16}-{313q^2 \over 924}+{4717Q_2 \over 3696}
\nonumber \\
&&~~~~~+{289 q^2 Q_2 \over 1848} -{38 Q_2^2 \over 77}
-{151Q_4 \over 176}+{17q_3 q \over 66} \biggr), \nonumber \\
&&a_{8,0}=M^9\biggl({35 \over 128}+{31303q^2 \over 64064}
-{101373 Q_2 \over 64064 } \nonumber \\
&&~~~~~~~~+{32171 q^4 \over 384384} -{22513 q^2 Q_2 \over 27456} 
+{57535 Q_2^2 \over 34944} \nonumber \\
&&~~~~~~~~~+{5571 Q_4 \over 4576}-{3343 q_3 q \over 6864} \biggr), \nonumber \\
&&a_{10,0}=M^{11}\biggl(-{63 \over 256}-{12974415 q^2 \over 20692672}
+{153744405  Q_2 \over 82770688 } \nonumber \\
&&~~~~~-{6074105  q^4 \over 22284416} +{1686890165  q^2 Q_2 \over 869092224}
-{2883979925 Q_2^2 \over 869092224} \nonumber \\
&&~~~~~-{1154583 Q_4 \over 739024} +{2221047 q_3 q \over 2956096} \biggr), 
\eeqn
and
\beqn
&&a_{2,1}={ 3i M^4 \over 2}\Bigl(-q+q_3\Bigr),\nonumber \\
&&a_{4,1}=iM^6\biggl(
{15q \over 8}+{15 q^3 \over 28}-{29q Q_2 \over 28}-{13 q_3\over 4}\biggr)
,\nonumber \\
&&a_{6,1}=i M^8\biggl( -{35q \over 16}-{3763 q^3 \over 2002}
+{95213q Q_2 \over 24024}+{30743 q_3\over 6864}\biggr) ,\nonumber \\
&&a_{8,1}=i M^{10}\biggl(
{315 q \over 128}+{579957 q^3 \over 155584}\nonumber \\
&&~~~~~~~~~~~~~~~~~~~~~
-{154403q Q_2 \over 19448}-{876263 q_3 \over 155584}\biggr). 
\eeqn  

\section{Formulas of $E$ and $\ell$ at the ISCO}

{}From Eqs. (\ref{eqene}) and (\ref{eqang}) with an approximate 
formula of $\rho_{\rm ISCO}$ shown in Sec. II, 
$E$ and $\ell$ at the ISCO are derived as 
\beqn
&&E_{\rm ISCO}=0.94281 - 0.03208 q -0.02975q^2 + 0.00794Q_2 \nonumber \\
&&~~~~~~~~~~-0.0341q^3 +0.0198qQ_2 -0.0019 q_3 \nonumber \\
&&~~~~~~~~~~- 0.0440q^4 +0.0404 q^2 Q_2 -0.0033 Q_2^2 \nonumber \\
&&~~~~~~~~~~+ 0.0005Q_4 - 0.0062q q_3, \label{eqa1}\\
&&{\ell_{\rm ISCO} \over M}=
3.4641 - 0.9428q -0.4444q^2 + 0.1879 Q_2 \nonumber \\
&&~~~~~~~~~~-0.3953 q^3+0.2996qQ_2-0.0392 q_3 \nonumber \\
&&~~~~~~~~~~-0.4470q^4 +0.4944 q^2 Q_2-0.0505Q_2^2\nonumber \\ 
&&~~~~~~~~~~+0.0093Q_4-0.0926 q q_3. \label{eqa2}
\eeqn
On the other hand, exact solutions for the Kerr case\cite{COMPACT} 
are expanded as 
\beqn
&&E_{\rm ISCO}^{\rm Kerr}=0.94281 - 0.03208 q -0.02182q^2 \nonumber \\
&&~~~~~~~~~~-0.01633q^3 - 0.01294q^4 \\ 
&&{\ell_{\rm ISCO}^{\rm Kerr} \over M}=
3.46410 - 0.94281q -0.25660q^2 \nonumber \\
&&~~~~~~~~~~-0.13531 q^3-0.08791q^4. 
\eeqn
Hence, as in the case of $\Omega_{\rm ISCO}$ and $R_{\rm ISCO}$, 
the error of coefficients 
is less than  $10^{-4}$ for $O(\epsilon^2)$ terms, $\sim 10^{-3}$ for 
$O(\epsilon^3)$ terms, and $\sim 10^{-2}$ for $O(\epsilon^4)$ terms. 

If accretion of matter occurs from the ISCO to the central rotating body, 
the value of $q$ of the central star increases as
\beq
\delta q={\mu \over M}\biggl({\ell_{\rm ISCO} \over M}
-2q E_{\rm ISCO}\biggr),
\eeq
where $\mu$ is the rest mass of the accreting matter. 
Using Eqs. (\ref{eqa1}) and (\ref{eqa2}), 
an approximate formula for $\delta q$ is given as 
\beqn
&&\delta q ={\mu \over M}\biggl(
3.4641 - 2.8284q -0.3802q^2 + 0.1879 Q_2 \nonumber \\
&&~~~~~~~~~~-0.3358 q^3 + 0.2837qQ_2-0.0392 q_3 \nonumber \\
&&~~~~~~~~~~-0.3788q^4 +0.4548 q^2 Q_2-0.0505Q_2^2\nonumber \\ 
&&~~~~~~~~~~+0.0093Q_4 -0.0887 q q_3\biggr). \label{eqa3}
\eeqn
Thus we obtain the following conclusions from Eqs. (\ref{eqa1}), 
(\ref{eqa2}), and (\ref{eqa3}):
If accretion of matter occurs from the ISCO, the effect of 
$Q_2$ are that 
(1) the maximum energy release efficiency of accreting matter 
slightly decreases, and that (2) the rate of increase in 
the angular momentum and 
$\delta q$ of central body is increased. 

\section{Field equations for computing rotating stars and 
the numerical method}

In this appendix, we show field equation for $\nu$, $\omega$, $B$, and 
$\zeta$ for obtaining axially symmetric rotating stars, and briefly 
mentioned our numerical method to solve them. 
The field equations are as follows \cite{BI}. 
\beqn
&&\nu_{,rr}+{2 \over r}\nu_{,r}
+{1 \over r^2}\{ \nu_{,yy}(1-y^2)-2y\nu_{,y} \} \nonumber \\
&&~~~~=4\pi e^{2\zeta} \rho_b [ 2h (u^0)^2 - \{ 1 +(2-\Gamma)\varepsilon \}
e^{-2\nu} ]\nonumber \\
&&~~~~~~~+{1 \over 2} e^{-4\nu}B^2 (1-y^2)
\{ r^2 \omega_{,r}^2 + \omega_{,y}^2(1-y^2) \} \nonumber \\
&&~~~~~~~-{B_{,r} \over B}\nu_{,r}-{1 \over Br^2}B_{,y}\nu_{,y}(1-y^2),
\label{feq1} \\
%%%%%%%%%%%%%%%%
&&\hat \omega_{,rr}+{4 \over r}\hat \omega_{,r}
+{1 \over r^2}\{\hat \omega_{,yy}(1-y^2)-4y\hat \omega_{,y} \} \nonumber \\
&&~~~~=-16\pi e^{2\zeta} \rho_b h (u^0)^2 (1 -\hat \omega) 
+\biggl( 4\nu_{,r}-{3B_{,r} \over B} \biggr)\hat \omega_{,r} \nonumber \\
&& \hskip 2.5cm 
+{1-y^2 \over r^2}\biggl(4\nu_{,y}-{3B_{,y} \over B}\biggr)\hat \omega_{,y},\\
%%%%%%%%%%%%%%%%
&&B_{,rr}+{3 \over r}B_{,r}+{1 \over r^2}\{ B_{,yy}(1-y^2)-3yB_{,y} \}
\nonumber \\
&&~~~~=16\pi B e^{2\zeta-2\nu}P,\\
%%%%%%%%%%%%%%%
&&\xi_{,y}=[(B+B_{,r}r)^2(1-y^2)+\{By-B_{,y}(1-y^2)\}^2]^{-1}
\nonumber \\
&&~~~~[(B+B_{,r}r)\{ -e^{-4\nu}\omega_{,y}\omega_{,r}r^3B^3(1-y^2)^2/2
\nonumber \\
&&~~~~~~+rB_{,ry}(1-y^2)-B_{,r}ry+2\nu_{,y}\nu_{,r}rB(1-y^2) \} \nonumber \\
&&~~~~~~-{1 \over 2}
\{By-B_{,y}(1-y^2)\} \{ e^{-4\nu}B^3r^2(1-y^2)(\omega_{,r}^2r^2 \nonumber \\
&&~~~~~~-\omega_{,y}^2(1-y^2))/2-B_{,rr}r^2-B_{,r}r-2B\nu_{,r}^2r^2
\nonumber \\
&&~~~~~~+B_{,yy}(1-y^2)-3yB_{,y}+2\nu_{,y}^2(1-y^2)B \}], \label{feq4}
\eeqn
where instead of $\omega$, 
we introduce $\hat \omega \equiv \omega \Omega_s^{-1}$ 
because $\omega \propto \Omega_s$ for slowly rotating cases. 

We solve Poisson type equations for $\nu$, $\omega~(\hat \omega)$, 
and $B$ as the boundary value problem. That is, we solve 
these equations imposing boundary conditions 
(\ref{boundM})$-$(\ref{boundB}) at $r=r_{\rm max}> 10M$, regularity 
condition at $r=0$ and $\theta=0$, and reflection 
symmetry conditions at $\theta=\pi/2$. 
On the other hand, the 
equation for $\zeta$ is solved using second order Runge-Kutta method 
imposing the boundary condition at $\theta=0$ as $\zeta= \ln B$ \cite{BI}. 

Using field variables, Eq. (\ref{beleq}) is rewritten as 
\beqn
&&(1+n K\rho_b^{1/n})^2
[-e^{2\nu}+\Omega_s^2(1-\hat \omega )^2 B^2 e^{-2\nu} r^2\sin^2\theta ]
\nonumber \\
&&=C (={\rm constant}). \label{feq5}
\eeqn
Here, $C$, $K$, and $\Omega_s$ are constants determined in iteration 
which is carried out in the following manner \cite{KEH}: 

\begin{enumerate}
\item First of all, we fix the coordinate radii of stellar surfaces at 
equator and pole. Also, we fix the value of $\rho_b$ at $r=0$. 
\item We give a trial density configuration for $\rho_b$. 
\item We solve field equations (\ref{feq1})$-$(\ref{feq4}). 
\item Since we have constraints imposed at procedure 1, 
we can determine $C$, $K$, and $\Omega_s$ using Eq. (\ref{feq5}).  
\item Using Eq. (\ref{feq5}) with 
$C$, $K$, and $\Omega_s$ determined at procedure 4, 
we calculate a new trial configuration for $\rho_b$. 
\item Return to 2, and repeat procedures 2$-$5 until 
a sufficient convergence is achived. 
\end{enumerate}

Numerical computation is typically performed taking 
$(N_r,N_{\theta})=(800,80)$ grid points, where 
$N_r$ and $N_{\theta}$ denote grid points for $0<r \leq r_{\rm max}$ and 
$0<\theta<\pi/2$. We use homogeneous grids in $r$ and $y=\cos\theta$. 
We check our numerical code by comparing our numerical results 
for non-dimensional quantities $MK^{-n/2}$, $JK^{-n}$, $RK^{-n/2}$, 
and $\Omega_sK^{n/2}$ 
with another results presented previously 
in ref.\cite{CST} for the case $n=1$. 
We find that our numerical results agree well with those in 
ref.\cite{CST} (inconsistency is less than $1\%$).

\clearpage 

{\bf Table I.~} Numerical data for $n=1$ polytrope which we 
adopt in Sec. III. Note that 
$R_e$ denotes the circumferential radius at the stellar surface on 
the equatorial plane. Because $M\Omega_s(\equiv \chi)$ 
is an invariant quantity irrespective of scaling of $K$, 
we can calculate the spin angular frequency of stars 
from the following data approximately as 
$370{\rm Hz} (\chi/0.016)(1.4M_{\odot}/M)$. 

\vskip 5mm
\begin{center}
\begin{tabular}{|c|c|c|c|c|c|c|c|c|} \hline
~$\rho_{b,0}K^n$~~ & $MK^{-n/2}$ & $R_{e}K^{-n/2}$ &~$\Omega_s K^{n/2}$~ 
& ~~~~~~~$q$~~~~~~~ & ~~~~~~~$Q_2$~~~~~~~ &~~~~~ $q_3$~~~~~ & 
$R_{\rm ISCO}/6M$ &$\Omega_{\rm ISCO}6\sqrt{6}M$\\ \hline
    .200 &    .158 &    .870 &    .061 &  9.48E-02&  3.53E-02 &  6.9E-03 &     .953 &    1.068 \\ \hline
    .217 &    .160 &    .852 &    .064 &  9.43E-02&  3.19E-02 &  6.1E-03 &     .953 &    1.069 \\ \hline
    .236 &    .162 &    .833 &    .067 &  9.38E-02&  2.91E-02 &  5.4E-03 &     .953 &    1.069 \\ \hline
    .257 &    .163 &    .814 &    .070 &  9.34E-02&  2.66E-02 &  4.8E-03 &     .952 &    1.070 \\ \hline
    .280 &    .164 &    .795 &    .073 &  9.31E-02&  2.41E-02 &  4.4E-03 &     .952 &    1.070 \\ \hline
    .305 &    .164 &    .776 &    .076 &  9.28E-02&  2.22E-02 &  4.0E-03 &     .952 &    1.070 \\ \hline
    .206 &    .160 &    .869 &    .098 &  1.50E-01&  8.45E-02 &  2.6E-02 &     .931 &    1.105 \\ \hline
    .224 &    .162 &    .851 &    .103 &  1.49E-01&  7.66E-02 &  2.3E-02 &     .930 &    1.106 \\ \hline
    .243 &    .163 &    .832 &    .107 &  1.49E-01&  6.99E-02 &  2.1E-02 &     .929 &    1.108 \\ \hline
    .265 &    .164 &    .813 &    .112 &  1.48E-01&  6.36E-02 &  1.8E-02 &     .928 &    1.109 \\ \hline
    .289 &    .165 &    .794 &    .116 &  1.47E-01&  5.82E-02 &  1.7E-02 &     .927 &    1.110 \\ \hline
    .314 &    .165 &    .775 &    .121 &  1.47E-01&  5.36E-02 &  1.5E-02 &     .927 &    1.111 \\ \hline
    .222 &    .163 &    .863 &    .144 &  2.12E-01&  1.53E-01 &  6.6E-02 &     .908 &    1.145 \\ \hline
    .241 &    .165 &    .844 &    .150 &  2.11E-01&  1.39E-01 &  5.9E-02 &     .906 &    1.149 \\ \hline
    .262 &    .166 &    .825 &    .156 &  2.10E-01&  1.27E-01 &  5.3E-02 &     .904 &    1.152 \\ \hline
    .285 &    .166 &    .806 &    .163 &  2.10E-01&  1.17E-01 &  4.8E-02 &     .902 &    1.154 \\ \hline
    .310 &    .167 &    .787 &    .170 &  2.09E-01&  1.07E-01 &  4.3E-02 &     .901 &    1.157 \\ \hline
    .237 &    .166 &    .859 &    .181 &  2.60E-01&  2.11E-01 &  1.1E-01 &     .891 &    1.176 \\ \hline
    .257 &    .167 &    .840 &    .189 &  2.59E-01&  1.93E-01 &  9.9E-02 &     .888 &    1.181 \\ \hline
    .280 &    .168 &    .820 &    .197 &  2.58E-01&  1.77E-01 &  8.9E-02 &     .885 &    1.186 \\ \hline
    .304 &    .168 &    .801 &    .205 &  2.57E-01&  1.63E-01 &  8.1E-02 &     .883 &    1.190 \\ \hline
    .231 &    .167 &    .876 &    .206 &  3.03E-01&  2.86E-01 &  1.7E-01 &     .882 &    1.196 \\ \hline
    .251 &    .168 &    .857 &    .214 &  3.01E-01&  2.61E-01 &  1.6E-01 &     .877 &    1.203 \\ \hline
    .272 &    .169 &    .838 &    .223 &  3.00E-01&  2.40E-01 &  1.4E-01 &     .873 &    1.210 \\ \hline
    .295 &    .170 &    .818 &    .232 &  2.99E-01&  2.21E-01 &  1.3E-01 &     .870 &    1.216 \\ \hline
    .249 &    .172 &    .883 &    .258 &  3.71E-01&  3.84E-01 &  2.9E-01 &     .862 &    1.237 \\ \hline
    .270 &    .173 &    .863 &    .269 &  3.70E-01&  3.54E-01 &  2.6E-01 &     .856 &    1.247 \\ \hline
    .292 &    .173 &    .844 &    .279 &  3.68E-01&  3.27E-01 &  2.4E-01 &     .851 &    1.257 \\ \hline
    .274 &    .176 &    .887 &    .307 &  4.28E-01&  4.53E-01 &  3.9E-01 &     .844 &    1.276 \\ \hline
    .294 &    .177 &    .869 &    .318 &  4.26E-01&  4.22E-01 &  3.5E-01 &     .837 &    1.288 \\ \hline
\end{tabular}
\end{center}

\clearpage

\begin{figure}[t]
\epsfxsize=2.8in
\begin{center}
\leavevmode
\epsffile{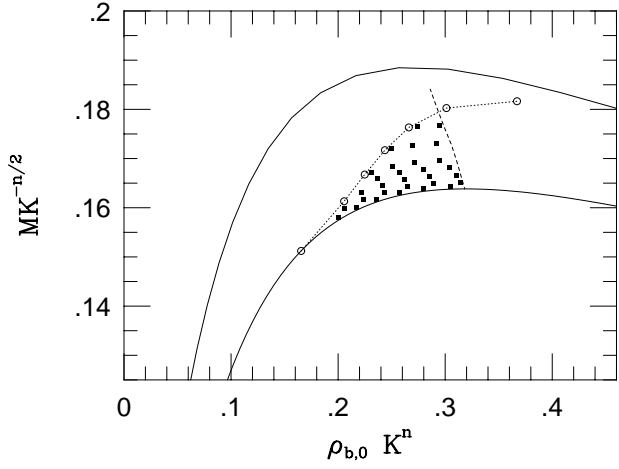}
\end{center}
\caption{Relation between mass ($MK^{-n/2}$) and density at $r=0$ 
($\rho_{b,0}K^n$) for n=1 polytrope. 
Lower and upper solid lines denote 
the relations for the spherical star and for the rotating star 
at the mass shed limit. 
Dotted line (with open circles) divides 
two regions where the ISCO exists or not: It exists below the line.
Dashed line divides the stable and unstable branches:
The left-hand side is the stable region. 
Filled squares denote data sets which are compared to our analytic
formulas.  
}
\end{figure}
\break

\begin{figure}[t]
\epsfxsize=2.8in
\begin{center}
\leavevmode
\epsffile{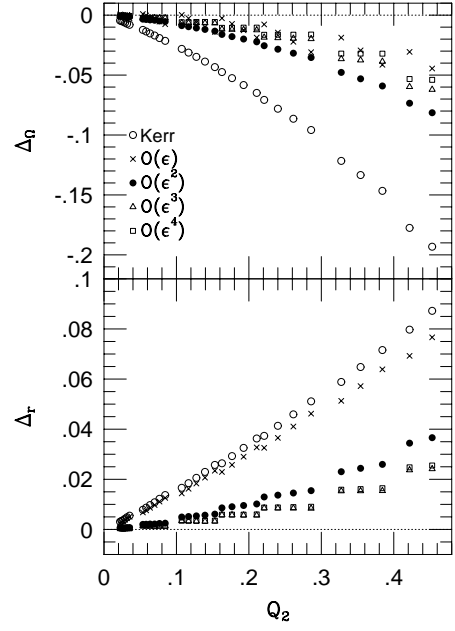}
\end{center}
\caption{$\Delta_{\Omega}$ and $\Delta_r$ as functions of $Q_2$ for $n=1$ 
polytrope. Open circles, crosses, filled circles, open triangles and 
open squares denote 
$\Delta_{a}^{\rm Kerr}$, $\Delta_{a}(1)$, 
$\Delta_{a}(2)$, $\Delta_{a}(3)$ and 
$\Delta_{a}(4)$~($a=\Omega~{\rm or}~r$), 
respectively. }
\end{figure}
%\clearpage

\end{document}